\documentclass[aps,prl,twocolumn,showpacs,groupedaddress]{revtex4}
\usepackage{graphicx}
\begin{document}
\title{Excitation spectroscopy of vortex lattices in a rotating 
Bose-Einstein condensate}
\author{S. J. Woo, L. O. Baksmaty, S. Choi, and N. P. Bigelow}

\affiliation{Department of Physics and Astronomy, and Laboratory for Laser
Energetics, University of Rochester, Rochester, NY 14627}
%\address{Department of Physics and Astronomy, University of Rochester, Rochester, NY 
%14627}

\date{\today}

\begin{abstract}
Excitation spectroscopy of vortex lattices in rotating Bose-Einstein 
condensates is described.  We numerically obtain the Bogoliubov-deGenne 
quasiparticle excitations for a broad range of energies and analyze them in 
the context of the complex dynamics of the system. 
Our work is carried out in a regime in which standard hydrodynamic assumptions 
do not hold, and includes features not readily contained within existing 
treatments.
\end{abstract} 
\maketitle 

\vspace{6mm}
The collective excitations of many particle systems play a central role in the understanding of the system's bulk properties and provide a basis for interpreting complex microscopic dynamics. In the case of a ``super"-system - a superfluid or a superconductor - studies of excitations can be used to probe particle-particle interactions and afford essential insight into energy flow and dissipation. Furthermore, the novel topological character of excitations such as vorticies are of intrinsic fundamental interest.  Stimulated by the achievement of vortex lattices in trapped atomic Bose-Einstein condensate (BEC) 
\cite{Ketterle, JILA}, there is an intense effort to understand quantized  vortices and their excitations in this system \cite{FetterReview,Dalibard_Hodby}.  Although vortex lattices have been extensively studied in superfluid $^4$He and type-II superconductors, most treatments rely on approximations that do not adequately reflect the situation of the trapped atomic BEC. 
For example, in the atomic BEC, the motion of the vorticies (and vortex lines) cannot be completely isolated from the bulk 
motion of the fluid in which they are imbedded.  To address these limitations, a number of studies have been presented \cite{ButtsRokhsar_FetterLattice_Tsubota_Kasamatsu,CastinDum}, 
including treatments of Tkachenko oscillations \cite{Anglin_Baym} 
and hydrodynamic shape oscillations \cite{StringariLattice,paper3}.
Nevertheless, a full analytical solution, valid over a wide range of energies and densities, is difficult if not impossible to construct due to the inherent complexity of the system.  In this paper we describe a comprehensive approach based on a numerical calculation.  We concentrate on the regime in which various 
hydrodynamic assumptions, such as the rigid body approximation, break down.
Our method enables us to search for new effects and to consider the case of a small condensate with few vortices of varying core size.

We first present a general description of the normal modes based on the observed structure of the quasiparticles.  We then discuss the excitation spectrum.  We find that the absence of the rigid body rotation of the fluid modifies the symmetry of the energy spectrum.  We also find that the spectrum is decorated by features arising from resonant processes that occur when wavelength of the surface excitations matches the intervortex distance.

Since we are interested in the rotating condensate, a
rotating reference frame is introduced and so the grand canonical
free Hamiltonian $\hat{F}$ includes additional centrifugal term:
\begin{eqnarray}
\label{free energy}
\hat{F}=\int d{\bf r}\ 
        \hat{\psi}^\dagger\left({\cal H}-\mu-\Omega{\cal L}_z\right)\hat{\psi}
      + \frac{1}{2} g\hat{\psi}^\dagger\hat{\psi}^\dagger\hat{\psi}\hat{\psi},
%&=&\hat{H}-\mu\hat{N}-\int d{\bf r}\ \hat{\psi}^\dagger \left(\Omega{\cal L}_z\right)\hat{\psi}
\end{eqnarray}
where ${\cal H}=-(\hbar^2/2M)\nabla^2+V_{\rm tr}({\bf r}), 
{\cal L}_z=-i\hbar\partial/\partial\phi$,
$\Omega$ is the rotation frequency of the reference frame,
$g = 4\pi a\hbar^2/M$ is the interparticle coupling constant with the 
$s$-wave scattering length $a$, $\hat{\psi}$ is a field operator satisfying
the bosonic commutation relation and $V_{\rm tr}({\bf r})$ 
is the trapping potential. 
We use harmonic trapping potential 
$V_{\rm tr}=M\omega_{\rm tr}^2(x^2+y^2)/2+M\omega_z^2z^2/2$.
Using the standard procedure in which the mean field 
$\psi({\bf r})\equiv\left<\right.\hat{\psi}({\bf r},t)\left.\right>$
and the fluctuation 
$\hat{\phi}({\bf r},t)\equiv\hat{\psi}({\bf r},t)-\psi({\bf r})$
are defined, the Gross-Pitaevskii equation (GPE) in the rotating frame at 
$T=0$ is given from the free energy [Eq. (\ref{free energy})] to be
\begin{equation}
\label{GPeqn_r11}
\left({\cal H}+g\psi^*\psi-\mu -\Omega{\cal L}_z\right)\psi =0,
\end{equation}
while applying the Bogoliubov transformation,
$\hat\phi\left({\bf r},t\right) = 
\sum u_j({\bf r})\hat{\alpha_j}e^{-i\omega_jt}
-{v_j}^*({\bf r}){\hat{\alpha_j}}^\dagger e^{i\omega_jt}$,
yields coupled Bogoliubov equations in the 
rotating frame,
\begin{eqnarray}
\nonumber
\left({\cal H}+2g\psi^*\psi -\mu -\Omega{\cal L}_z\right)u_j-g\psi^2v_j&=&\hbar\omega_ju_j \\
[-.2cm]
\label{bogol_r11} \\
[-.2cm]
\nonumber
\left({\cal H}+2g\psi^*\psi -\mu +\Omega{\cal L}_z\right)v_j-g\psi^{*2}u_j&=&-\hbar\omega_jv_j. 
\end{eqnarray}

These equations, written as sparse matrices of order $10^5\times 10^5$, are 
solved numerically using parallel computing packages \cite{paper2}.
All calculations are done in a quasi 2-dimensions (2D) limit consistent with 
the experimentally relevant pancake geometry\cite{JILA}. 
All values given in this paper are based on the dimensions of the trap potential: 
$\sqrt{\hbar/M\omega_{\rm tr}}$ for the length and 
$\omega_{\rm tr}^{-1}$ for time. 
For these 2D calculations, the effective coupling constant $g_{\rm 2D}$ is used, which is 
defined $g_{\rm 2D}=g(m\omega_z/2\pi\hbar)^{1/2}$ \cite{CastinDum}.
We use $\omega_{\rm tr}=2\pi\times 74.25{\rm Hz}$, $\omega_z=10\omega_{\rm tr}$ and 
$10^4$ rubidium atoms with $a=5.77{\rm nm}$ which gives $g_{\rm 2D}=0.0724$ in 
trap potential dimensions.

To calculate the ground state we propagate an initial trial state in imaginary 
time, a technique which consistently decreases the energy until convergence is observed.
For the above parameters, a triangular vortex lattice state containing 31 vortices with a six-fold
symmetry is found for a trap rotation frequency $\Omega = 0.8$, 
while one with 7 vortices with the same symmetry is found for
$\Omega = 0.5$. 
The lattice with 31 vortices has a Thomas-Fermi (TF) radius $r_{\rm TF}=7.1$, 
a healing length $\xi=0.23$ with an intervortex spacing $b=2.1$;  
corresponding values for the lattice with 7 vortices are $r_{\rm TF}=5.9$, $\xi=0.19$ 
and $b=2.7$. 
It is clear from the relatively large ratio $\xi/b$ that these systems are 
not strictly in a hydrodynamic regime and, especially for the case of 7 vortices, that
the large $b/r_{TF}$ ratio indicates that the system does not have sufficient number of 
vortices to ensure a rigid body rotation of the fluid.

We can visualize the evolution of the density profile of the $j$th normal mode using the
time-dependent probability density $|\psi + u_{j}e^{-i\omega_{j} t} -
v_{j}^{*}e^{i \omega_{j} t}|^{2}$, where $\psi({\bf r})$ indicates the ground
state wave function and $\omega_{j}$ represents the corresponding eigenenergy. 
Examining the time evolution, we observe that the Tkachenko modes
occur only in the lower energy regime.  The time dependent density profiles for the 
Tkachenko modes were found to be in excellent agreement with the experimental results from JILA \cite{JILA} and are described elsewhere\cite{paper2}.  We also find that the hydrodynamic shape excitations which 
involve significant motion of the bulk fluid are generally found throughout the 
entire energy range.

Following \cite{Fliesser_Ohberg}, we note that in the absence of the lattice, hydrodynamic bulk excitations may be classified by the radial and angular quantum 
numbers $n$ and $m$. 
In the presence of the vortex lattice we seek a similar description. First, we find that the amplitudes for 
$u$ and $v$, when plotted as a function of position, typically show an
annular symmetry, with density dips at the position of the individual vortices.
In the radial direction, the nodes in $u$ and in $v$ are quite 
clear and the number of nodes can be used to define the radial quantum numbers $n$.  
We next observe that the phase of $u$ is no longer the same as that of $v$. 
Specifically, we find that if we write $\psi=\psi_0e^{iS_0}$ then $u$
and $v$ are $u_0e^{iS+iS_0}$ and $v_0e^{iS-iS_0}$ respectively, 
where $\psi_0$, $u_0$, $v_0$, $S$ and $S_0$ are real functions of ${\bf r}$;  
$n^{\prime}\equiv \psi^* u - \psi v$ \cite{FetterReview} is therefore adopted
to characterize the phase properties of the excitations.
Phase plots of $n'$ generally indicate that a well-defined multiple 
of $2\pi$-phase change occurs around the edge of the cloud, even though the angular dependence of 
$n^{\prime}$ is often more complicated than $e^{im\phi}$. 
Mathematically, $m$ may be defined by a contour integration of $\nabla S$ 
along a closed contour enclosing the condensate at the Thomas-Fermi 
radius: 
$m=\frac{1}{2\pi}\oint_{\rm c}d{\bf r}\cdot\nabla S $.
Temporal evolution of hydrodynamic shape modes have shown that they correspond well
to circulating (annular) waves with $|m|$ wavelengths and that the sign of $m$ determines 
the direction of the rotating wave;
they rotate counter-clockwise if $m>0$, and clockwise if $m<0$.

\begin{figure}[ht]
\centerline{(a) \hspace{3.8cm} (b)}
\centerline{\includegraphics*[height=2.5cm]{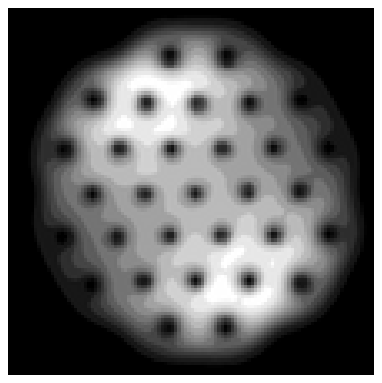}\hspace{2cm}
            \includegraphics*[height=2.5cm]{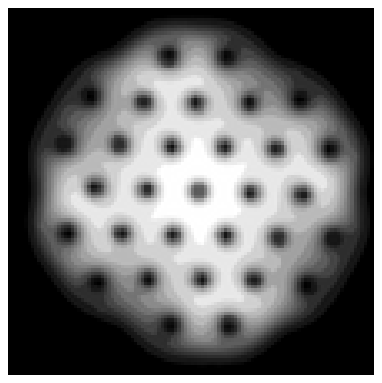}
           }
\vspace{0.01cm}
\centerline{(c) \hspace{3.8cm} (d)}
\centerline{\includegraphics*[height=2.5cm]{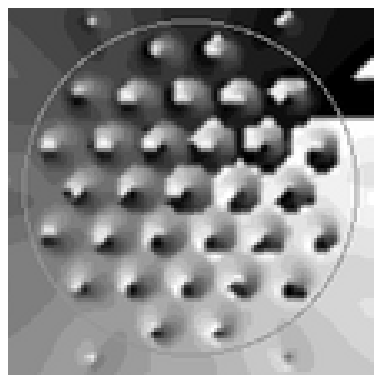}\hspace{2cm}
            \includegraphics*[height=2.5cm]{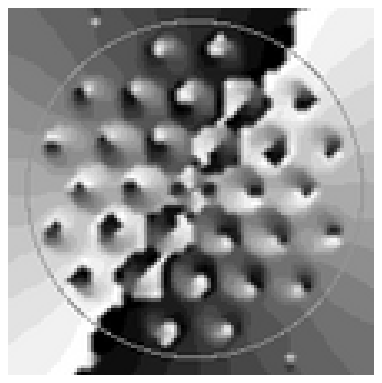}
           }
\caption{(a) Dipole mode ($n=0$, $m=1$) with $\omega=1-\Omega=0.20$ and
(b) quadrupole mode ($n=0$, $m=2$) with $\omega=0.36$.
Surface plot of phase from $0$ to $2\pi$ (dark to light) across $n'$ 
(c) for the dipole mode and
(d) for the quadrupole mode.
It is clear that the phase change along the circle drawn in each plot is $2\pi$ for $m=1$ 
and $4\pi$ for $m=2$.}
\label{surface_mode}
\end{figure}
The simplest examples of the hydrodynamic excitations are the dipole 
(or the ``center-of-mass'') modes ($n=0$, $m=\pm 1$) and the quadrupole modes 
($n=0$, $m=\pm 2$) \cite{Haljan}.  
The $m=+1$ dipole mode displays two peaks in condensate density which rotate at a frequency 
$\omega=1-\Omega$ while the vortices remain fixed.  Meanwhile for the $m=-1$ mode, the whole 
condensate ``sloshes'' around a small circle together with the vortex lattice 
(without any visible lattice deformation) and occurs at a frequency $\omega=1+\Omega$. 
In the lab frame, the frequencies for the two dipole modes 
are simply the trap frequency, which is $1$ in our dimensions. 
Figures \ref{surface_mode}(a) and (b) show an instantaneous density profile (``snapshot") taken from
the temporal evolution of the $m=+1$ dipole and the $m=+2$ quadrupole modes respectively.
Figures \ref{surface_mode}(c) and (d) depict the corresponding phases of the 
noncondensate fluctuation density $n'$. 
Along a circular contour drawn in each plot at Thomas-Fermi radius, an integral 
number of $2\pi$-phase-changes can be equated to $m$. 
The ($n=1,m=0$) and ($n=2,m=0$) breathing modes were also be identified and occurred
at $\omega=2$ and $\omega=2.7$ respectively (for $\Omega=0.8$).   
For the breathing mode, the vortices move sympathetically with the bulk fluid 
motion, and there is no difference between the frequencies in the rotating and 
nonrotating reference frames. 

\begin{figure}[ht]
\centerline{(a) \hspace{3.8cm} (b)}
\centerline{\includegraphics*[height=2.5cm]{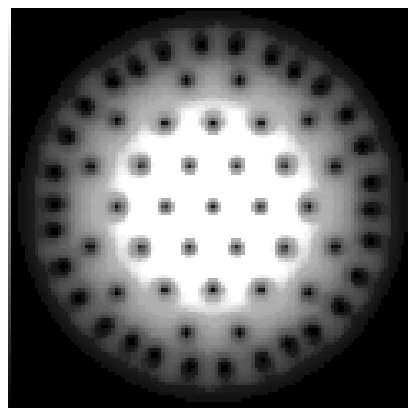}\hspace{2cm}
            \includegraphics*[height=2.5cm]{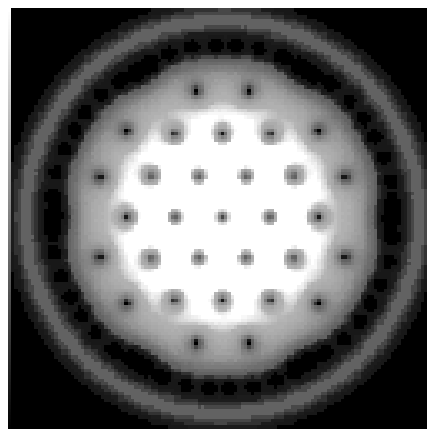}
           }
\vspace{0.01cm}
\caption{Density plot for $n=0$ and (a) $m=30$ (b) $m=50$ mode. 
(a) Surface wave forms visible vortices. (b) Vortices around condensate merge into 
each other.}
\label{vortex_sheet}
\end{figure}
We note additionally that a surface mode with quantum number $\pm m$ 
introduces $|m|$ nodes on the edge of the condensate, each of which involves 
a phase singularity with a $\pm 2\pi$-phase change 
around that singularity.
Although these nodes cannot be identified as vortices, particularly for small $m$, we find that as 
$m$ increase, we see $m$ vortices gradually forming.
We conjecture that this effect is linked to the vortex nucleation and 
annhilation processes, particularly
when a change of rotational speed is imposed on the system.
For sufficiently large $m$, the vortices merge with each other in a continuous manner 
and appear to form a vortex sheet (Fig. \ref{vortex_sheet}) \cite{Eltsov}.  
However, no clear evidence of the phase transition to vortex sheets has been
identified.

\begin{figure}[ht]
\centerline{(a) \hspace{3.8cm} (b)}
\centerline{\includegraphics*[height=4.3cm]{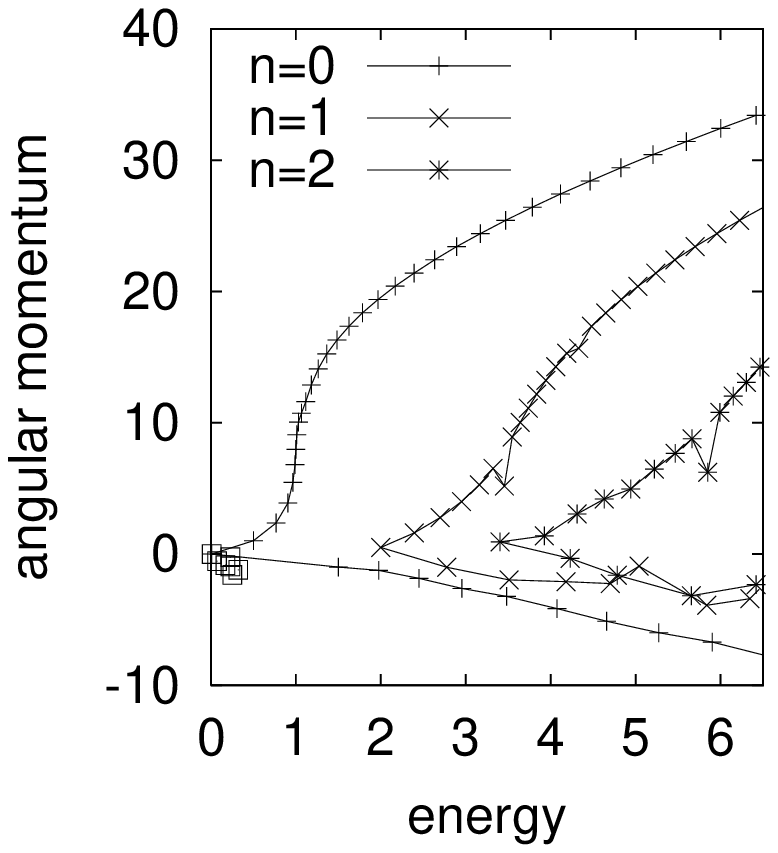}
            \includegraphics*[height=4.3cm]{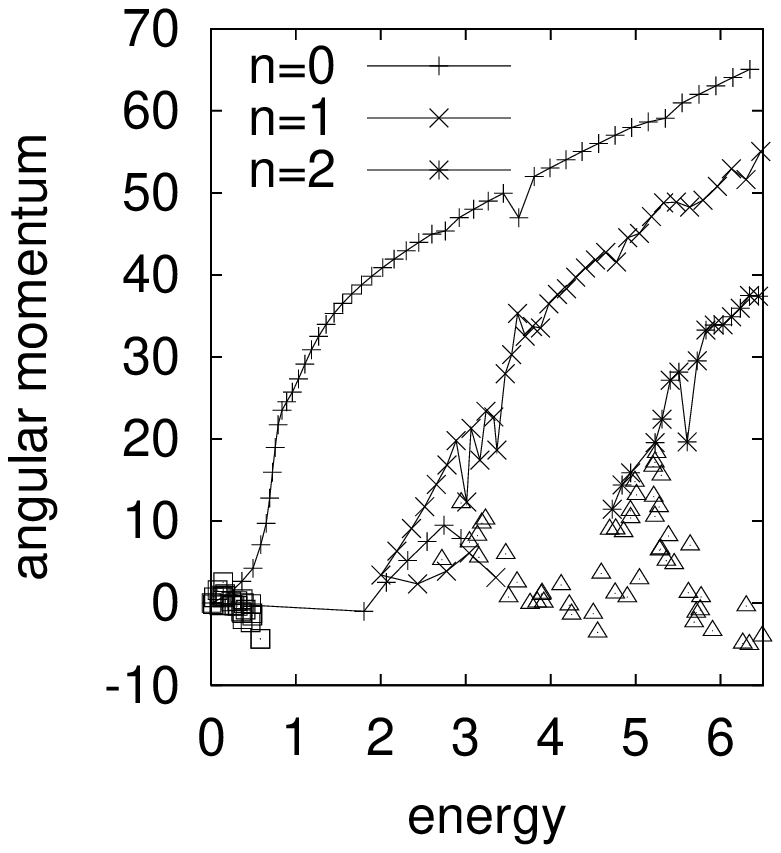}
           }
\vspace{0.01cm}
\centerline{(c) \hspace{3.8cm} (d)}
\centerline{\includegraphics*[height=4.3cm]{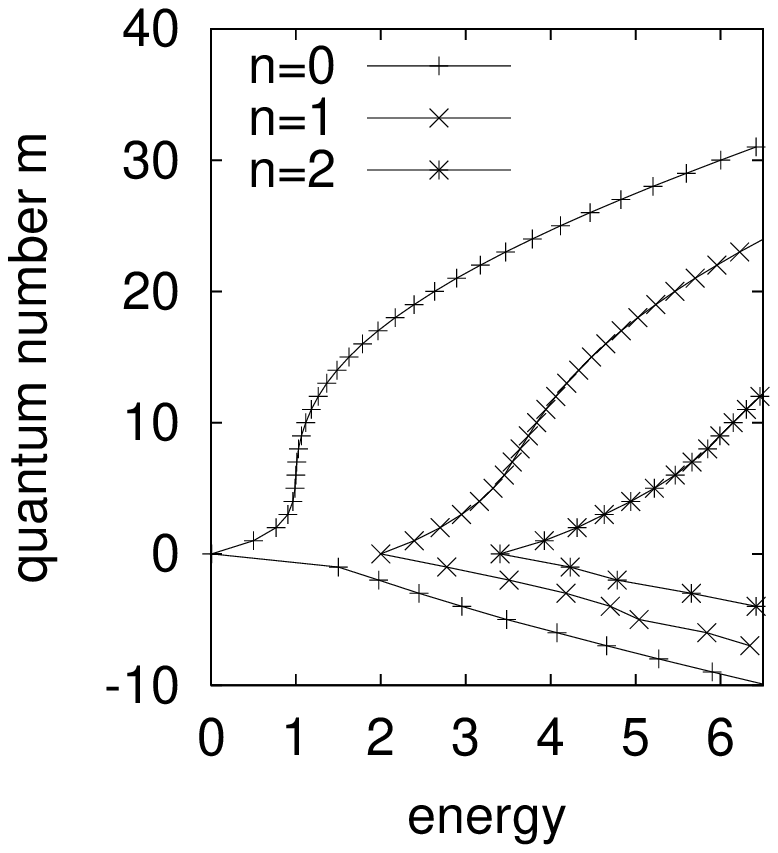}
            \includegraphics*[height=4.3cm]{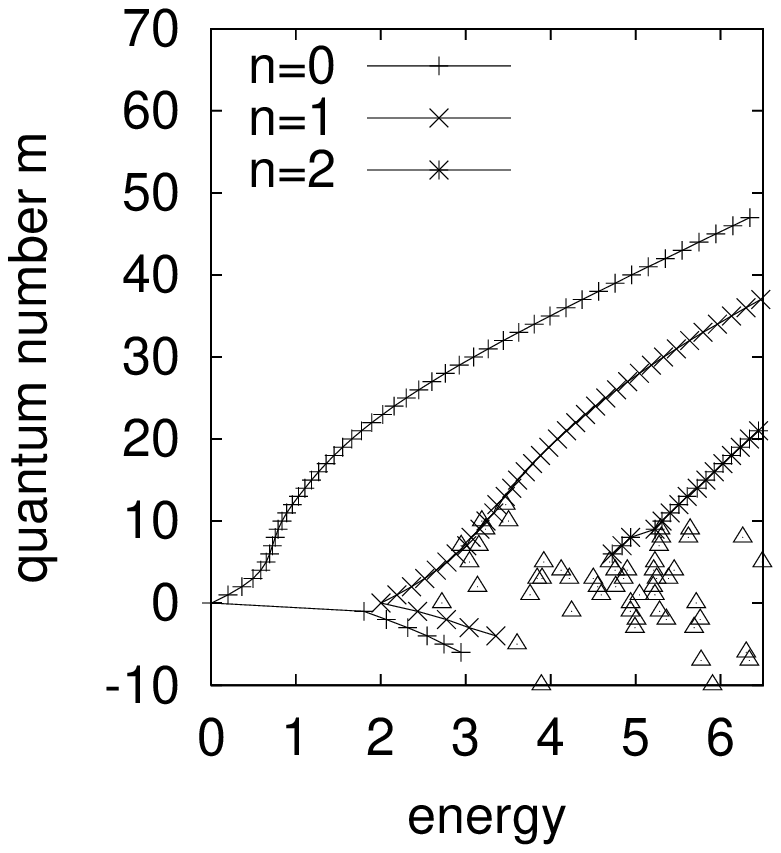}
           }
\caption{Plots of angular momentum $L_z$ {\it vs.} energy $\omega_{n,m}$ for 
(a) $\Omega = 0.5$ (7 vortices) and (b) $\Omega = 0.8$ (31 vortices), 
and angular quantum number $m$ {\it vs.} $\omega_{n,m}$ 
for (c) $\Omega=0.5$ and (d) $\Omega=0.8$.}
\label{dispersion}
\end{figure}

We construct the excitation spectrum in the form of a plot of the excitation mode angular momentum as a function of energy -- the central result of this paper. 
The angular momentum of the excitations with respect 
to the rotating condensate may be defined \cite{Isoshima} as 
$L_z=\left[(l_u-l_\psi)N_u+(l_v+l_\psi)N_v\right]/\left(N_u+N_v\right)$,
where 
$l_\alpha 
=\left(\int d{\bf r}\ \alpha^*(-i\hbar\partial/\partial\phi)\alpha\right)\left/
\left(\int d{\bf r}\ \alpha^*\alpha\right) \right.$
with $\alpha=u,v,\psi$ and $N_u=\int d{\bf r} |u|^2$, $N_v=\int d{\bf r} |v|^2$.
This gives the average difference in the angular momenta for the quasiparticles 
and the condensate ground state.
$L_z$ captures the effect of the detailed vortex lattice structure 
while $m$ is an effectively coarse-grain-averaged result.
The angular momentum $L_z$ {\it vs.} excitation energy $\omega_{n,m}$ curves 
for $\Omega=0.5$ (7 vortices) and $\Omega=0.8$ 
(31 vortices) and the corresponding $m$ {\it vs.} $\omega_{n,m}$ curves 
in the rotating frame are presented in Fig. \ref{dispersion}. 
The points with the same radial quantum number $n$ are connected 
such that different points on each line correspond to the same $n$ but to
different $m$. 
The leftmost curve corresponds to the surface mode with $n=0$, while the 
other curves are modes with higher $n$.  
In addition to these curves, we identify the Tkachenko modes as the group of 
points (the open squares) in the bounded region of very low energy.
The points which are not connected (open triangles) are those modes which display very strong coupling of the bulk fluid motion to the vortices, 
thus making exact identification of the mode character difficult. 
This happens particularly for small $m$ with higher $n$ modes, for which excitation is concentrated deep within the condensate.

In our previous work we found that, in the rotating frame, 
there is a small energy splitting $2\Omega$ between the $+m$ and 
$-m$ states independent of $|m|$, resulting in a spectrum which is highly symmetric about $m=0$ \cite{paper3}. 
This follows from the assumption that the vortex lattice causes the condensate to rotate as a rigid body. 
By contrast, the observed asymmetry in Fig. \ref{dispersion} is  
due to the non-rigid rotation effects:
If there are $N_{\rm vtx}$ vortices within the Thomas-Fermi radius 
$r_{\rm TF}$, the velocity of the condensate flow at $r_{\rm TF}$ is approximately
$v_{\rm TF}\approx 2\pi\hbar N_{\rm vtx}/Mr_{\rm TF}$
implying that the actual rotating frequency of the condensate at 
the Thomas-Fermi radius is
$\Omega_{\rm TF}=v_{\rm TF}/2\pi r_{\rm TF}=
       \hbar N_{\rm vtx}/Mr_{\rm TF}^2$.
In the case of 7 vortices, $r_{\rm TF}=5.9$, which 
yields $\Omega_{\rm TF}=0.2$.
The rotation of the condensate on the edge of the cloud is therefore much slower 
than the rotation of trap.
One can show that the frequency of the circulating wave in the reference frame
rotating at $\Omega_{\rm TF}$ is
$\omega_{n,m}'=\omega_{n,m}+m(\Omega-\Omega_{\rm TF})$ \cite{Footnote1} and if we use this 
$\omega_{n,m}'$ as the energy, then $m$ {\it vs.} $\omega_{n,m}'$ curve becomes
almost symmetric about $m=0$. 
With a sufficiently large number $N_{\rm vtx}$ (rigid body regime), 
$\Omega_{\rm TF}\approx\Omega$ \cite{FetterReview}.
In the case of 31 vortices, the energy splitting contribution proportional to $m$ became rather small 
while the predicted $2\Omega$ splitting persisted.

We note that for the positive $m$ states the $L_z$ {\it vs.} $\omega_{n,m}$ 
curve is quite smooth, yet there are several exciations with lower angular momenta 
that appear as ``dips'' in the curve (see Fig. \ref{dispersion}).  These dips 
occur at the same angular momentum even for different $n$.  
The quantum number $m$ for these states are approximately integral multiples of the 
number of outermost vortices, which is 6 for $\Omega = 0.5$ and 12 for 
$\Omega = 0.8$. 
This effect may be understood from the fact that the bulk excitations interact 
strongly with 
the vortex lattice when the intervortex spacing of the 
outermost vortices happens to be a multiple of the wavelength of the surface 
wave.
We have observed that for these states, vortices, resonant with the surface wave, 
partly share the excitation energy with the surface wave so as to reduce the angular momentum.  
This situation is reminiscent of the dispersion relation of electrons in an ion 
core lattice potential where, using a Kronnig-Penny model \cite{Kittel} one finds that deviations appears when an integral multiple electron wavefunction wavelengths coincides with the inter-ion spacing. 
The analogy is not complete, however, because the vortices can be more easily disturbed by the surface waves than can ion cores be displaced in a crystal.

\begin{figure}[ht]
\centerline{(a) \hspace{3.8cm} (b)}
\centerline{\includegraphics*[height=1.8cm]{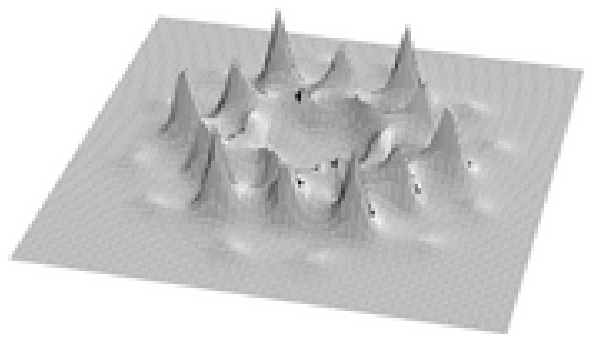}\hspace{1.5cm}
            \includegraphics*[height=1.8cm]{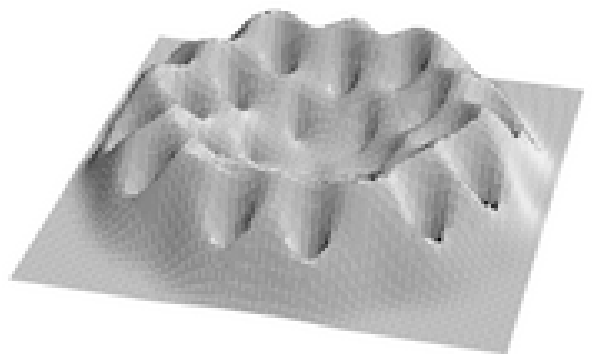}
           }
\centerline{(c) \hspace{3.8cm} (d)}
\centerline{\includegraphics*[height=1.8cm]{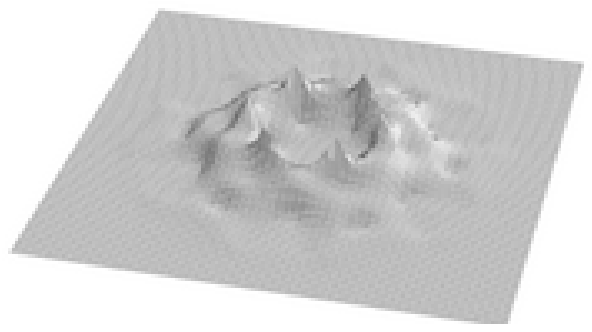}\hspace{1.5cm}
            \includegraphics*[height=1.8cm]{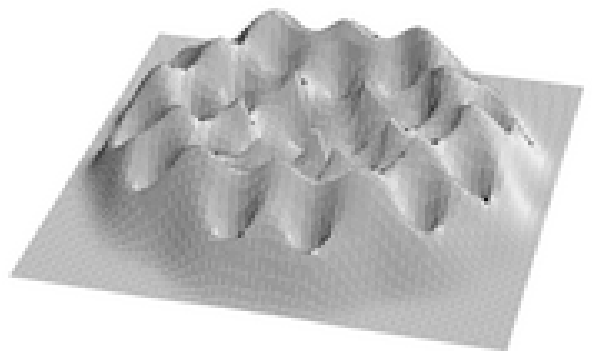}
           }
\vspace{0.01cm}
\centerline{(e)}
\centerline{\includegraphics*[height=1.8cm]{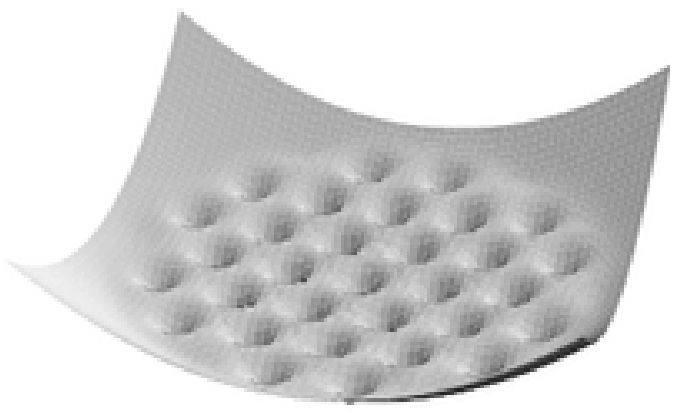}}
\caption{
Density plots of $u$ for (a) the Tkachenko and (b) the hydrodynamic ($n=0$, $m=1$) 
mode;
(c) and (d) are the corresponding plots of $v$. 
(e) gives the mean field potential that quasi particle $u$ and $v$ experience.}
\label{u_plot}
\end{figure}
Finally, to understand the Tkachenko and the 
hydrodynamic shape modes at the level of quasiparticles, we plot the density of 
$u$ and $v$ for these two cases. 
Figures \ref{u_plot}(a) and (b) show that, for the Tkachenko modes, 
$u$ and $v$ have localized peaks 
where the vortices are present, while for the hydrodynamic shape excitations, they are more 
evenly spread with depressions instead of peaks.  
The regions where the quasiparticles are localized are mainly excited.
In particular, vortex precession is observed for the Tkachenko modes. 
Figure \ref{u_plot}(c) shows the effective mean field potential the quasiparticles 
experience, $V_{\rm tr}+2g\psi^*\psi-\mu$, from the Bogoliubov
equation, Eq. (\ref{bogol_r11});
the vortex cores create an array of effective potential wells.
It is clear that the Tkachenko modes may be viewed as the lower energy bounded states 
within those wells.  
The hydrodynamic shape modes may be viewed as ``unbounded'' states with higher energies
 -- even though still bounded within the trapping potential.

In summary, we have described new aspects of normal modes of an atomic BEC supporting a vortex lattice. 
From a spectroscopic studies of the excitations, we found anomalous effects which happen due to the 
singularity of vorticity and due to the non-rigid rotation of the condensate. 
Some of our observations should be readily verifiable experimentally. 
Our results for the excitation spectrum 
is an ideal starting point for finite temperature calculations related 
to the dissipation and decay of vortices. This will be the subject of
future work.

During the preparation of the manuscript we learned of a related preprint \cite{Mizushima}.  This work is supported by the NSF, ONR and ARO. 
L. O. Baks. is a Horton Fellow.


\begin{references}
\bibitem{Footnote1} In the frame rotating at $\Omega$, 
the angular phase velocity of the normal mode with the excitation frequency 
$\omega_{n,m}$ is $\chi_{n,m}=\omega_{n,m}/m$.
Its angular phase velocity in the lab frame is 
$\chi_{n,m}^{\rm lab}=\chi_{n,m}+\Omega$. 
Assuming that the frequency of the same state in the frame rotating at 
$\Omega'$ is $\omega'_{n,m}$, $\chi'_{n,m}=\omega'_{n,m}/m$ and 
$\chi_{n,m}^{'\rm lab}=\chi'_{n,m}+\Omega'$. 
But $\chi_{n,m}^{\rm lab}=\chi_{n,m}^{'\rm lab}$ since they are for the 
same state in the lab frame, which yields 
$\omega'_{n,m}=\omega_{n,m}+m\left(\Omega-\Omega'\right)$
\end{references}
\end{document}